\documentclass[fleqn,10pt]{wlscirep}
\usepackage[utf8]{inputenc}
\usepackage[T1]{fontenc}
\usepackage{bm}
\title{Magnetodynamics of few nanoparticle chains}

\author[1,*]{Thinh Q. Bui}
\author[3,4]{Samuel D. Oberdick}
\author[2]{Frank M. Abel}
\author[5]{Michael J. Donahue}
\author[1]{Klaus N. Quelhas}
\author[2]{Cindi L. Dennis}
\author[2,6]{Thomas E. Cleveland IV}
\author[6,7]{Yanxin Liu}
\author[1]{Solomon I. Woods}
\affil[1]{Physical Measurement Laboratory, National Institute of Standards and Technology, Gaithersburg, MD, USA 20899}
\affil[2]{Material Measurement Laboratory, National Institute of Standards and Technology, Gaithersburg, MD, USA 20899}
\affil[3]{Physical Measurement Laboratory, National Institute of Standards and Technology, Boulder, CO, USA 80305}
\affil[4]{Department of Physics, University of Colorado, Boulder, CO, USA 20899}
\affil[5]{Information Technology Laboratory, National Institute of Standards and Technology, Gaithersburg, MD, USA 20899}
\affil[6]{Institute for Bioscience and Biotechnology Research (IBBR), Rockville, MD, USA 20850}
\affil[7]{Department of Chemistry and Biochemistry, University of Maryland, College Park, MD, USA 20742}
\affil[*]{e-mail: thinh.bui@nist.gov}

\begin{abstract}
In recent years, there has been increasing interest in the understanding and application of nanoparticle assemblies driven by external fields.  Although these systems can exhibit marked transitions in behavior compared to non-interacting counterparts, it has often proven challenging to connect their dynamics with underlying physical mechanisms or even to verifiably establish their structure under realistic experimental conditions.  We have studied colloidal iron oxide nanoparticles that assemble into ordered, few-particle linear chains under the influence of oscillating and pulsed magnetic fields.  In this work, our goal has been to answer the following question: by what physical mechanisms does the magnetic switching of a linear chain evolve from the switching of its constituent particles? Cryo-TEM has been used to flash freeze and image the structures formed by oscillatory drive fields, and magnetic relaxometry has been used to extract the multiple time constants associated with magnetic switching of the short chains.  Armed with the physical structure from microscopy and the field-dependent switching times from magnetic measurements, we have conducted extensive micromagnetic simulations, revealing probable physical mechanisms for each time constant regime spanning $10^{6}$ ($\approx$ 1 $\mu$s to 1 s) in time.  These types of magnetic nanomaterials have great potential for biomedical technologies, particularly magnetic particle imaging and hyperthermia, and rigorous elucidation of their physics will hasten their optimization.
\end{abstract}
\begin{document}

\flushbottom
\maketitle

\thispagestyle{empty}

\noindent \textbf{Key points:} magnetic nanoparticles, nanoscale assembly, magnetic particle imaging, magnetization dynamics, magnetometry

\subsection*{Introduction}

Below a critical size (ranging from $\approx$ 20 nm to $\approx$ 100 nm), ferromagnetic materials can form single domain magnetic nanoparticles (MNPs) that behave as macrospins --- all internal atomic spins are aligned and a single particle can be treated like a single dipole\cite{Butler_1975,Leslie-Pelecky_1996}. Interacting MNPs exhibit a rich variety of behavior including super spin-glass and superferromagnetism\cite{Jonsson_1998_2,Jonsson_1998,Morup_2010,Zelenakova_2014}. When magnetically-interacting MNPs are dispersed in liquid (ferrofluid) rather than solid matrix, the complexity of particle dynamics is even greater, as the magnetic moment can switch by internal spin dynamics (Néel mode - reorientation of magnetic moment inside a particle) and by physical rotation (Brownian mode - rotational diffusion of the whole particle in a fluid) \cite{Eberbeck_2006,Berkov_2006,Deissler_2014,Ota_2019}.  Furthermore, MNPs can assemble under fields through translational and rotational motion in the fluid, leading to dynamic change of inter-particle interactions \cite{Faraudo_2016,Anderson_2021}. The prospect of controllable and reversible assembly is indeed intriguing, especially when utilizing MNPs as functional constituents for artificial materials and devices (e.g. MRI contrast agents, drug delivery system, nano-/micro-scale robotics, metamaterials) \cite{Nie_2010, Bao_2016}. Dynamics of ferrofluids can be complex so that experimental and computational studies often consider Néel and Brownian mechanisms in isolation, but in a few cases the more realistic situation where both can operate simultaneously has been considered \cite{Berkov_2006,Weizenecker_2018,Ota_2019,Draack_2019}.  

Assemblies of MNP-based colloids have been well documented in MNPs ranging from nanometers to micrometers in size: see review by Li \textit{et al.}\cite{Li_2022}. In contrast to MNP assemblies formed under static or quasi-static conditions (DC or low frequency fields)\cite{Graeser_2007, Wang_2009, Butter_2003}, there has been recent and growing interest in understanding the dynamic magnetization of superparamagnetic nanoparticles and ensembles (chains or clusters) in AC excitation fields in the kHz to MHz range for applications of biomedical imaging, biosensing, and hyperthermia\cite{Fung_2023, Chugh_2023, Spatafora-Salazar_2021,Colson_2022}. Specifically, recent studies on the AC field-induced dynamic chain formation revealed a > 10-fold enhancement in both imaging resolution and sensitivity for magnetic particle imaging (MPI)\cite{Tay_2021,Colson_2022,Fung_2023}. The "dynamic" nature of chain formation refers to the time-varying dipolar interactions by using time-varying fields\cite{Spatafora-Salazar_2021}. However, in recent work\cite{Tay_2021} the physical picture of chains was obtained by TEM under static DC fields and dried on a TEM grid, which may obscure the actual structures under realistic (colloidal) conditions where chains are formed. 

Our work is motivated by the need for stronger evidence of the native physical structure and detailed physical mechanisms for chain dynamics.  In this pursuit, we have studied  ferrofluids comprised of MNPs with strong magnetic dipole interactions that undergo field-driven assembly, where the dipolar interactions range from weak (low field, single particles) to strong (high field, 1D particle chains). We first present AC magnetometry results to show distinctly different behavior compared to particles with relatively weaker interactions. When driven with an AC field, these strongly interacting dispersions show a remanent magnetization and rapid switching of magnetization at coercive fields. These observations suggest that particles form field-driven assemblies that can stabilize magnetic correlations between several particles\cite{Morales_2023}. 

To provide strong evidence for the multi-particle structure under the magnetic field conditions used for magnetic particle imaging, we performed experiments where particles were flash-frozen while driven by AC magnetic fields. Distinct from more complex chain assemblies in ferrofluids reported previously\cite{Butter_2003,Lalatonne_2004,Taheri_2015, Chen_2023}, our system is particularly simple, where the linear chains are relatively short (most of them have 3 to 5 particles) and magnetic interactions between particles are dominated by dipolar forces because the MNPs are coated by a thin non-magnetic shell (oleic acid/oleylamine). Cryo-TEM imaging has been used to capture the geometry of these structures in their native colloidal state, which validates the existence of short chains responsible for enhanced magnetic response. These observed structures provide essential information (chain length, inter-particle distance) for micromagnetic simulations of complex chain formation dynamics.

To reveal detailed mechanisms of chain formation and decay, we used magnetic measurements and numerical simulations, constrained by the physical structure determined by cryo-TEM. The magnetodynamics of these short chains are probed with custom magnetic instrumentation \cite{Tay_2016,Bui_2022,Fung_2023} to isolate the separate Néel and Brownian dynamics involved in switching of the chains over timescales from 1 $\mu s$ to 10 s. Then, using a combination of energetics arguments and micromagnetic simulations based on the Landau–Lifshitz–Gilbert (LLG) equation computed using finite difference methods from the OOMMF software package\cite{Palacios_1998,Donahue_1999}, we show that chain structures with only a few particles can explain our observations from AC magnetometry. Our linear chains exhibit hysteresis in the magnetic moment ($m$) vs. applied magnetic field ($H$) data and a multi-step relaxation process that is analyzed to extract time constants for each step. Direct comparison of the magnetic data with micromagnetic simulations enables us to discriminate between Néel and Brownian processes and connect each step in the relaxation process with different physical mechanisms. Finally, we connect the measured time constants and switching amplitudes to the different magnetodynamic mechanisms that can be directly correlated with findings from micromagnetic simulations. 

\subsection*{Magnetometry of Weakly Interacting (WI) and Strongly Interacting (SI) MNPs}

Colloidal dispersions of MNPs can show markedly different responses to AC magnetic fields, depending on the strength of dipolar interactions between constituent particles. In weakly interacting (WI) MNPs, the magnetization shows low remanence and coercivity\cite{Kaiser_1970, Hu_2019}, similar to a superparamagnetic response. In strongly interacting dispersions, however, the enhanced magnetization can be stabilized through dipolar interactions\cite{Morup_2010}. The magnetic correlations between particles can persist on timescales similar to the period of the AC field and longer, leading to observable coercivity and remanence (this behavior is sometimes referred to as superferromagnetism in the literature\cite{Tay_2021,Fung_2023,Colson_2022}). Throughout this work, we use the terms SI and WI to distinguish between strongly interacting and weakly interacting dispersions, respectively.

\begin{figure}
     \centering
     \includegraphics[width=15cm]{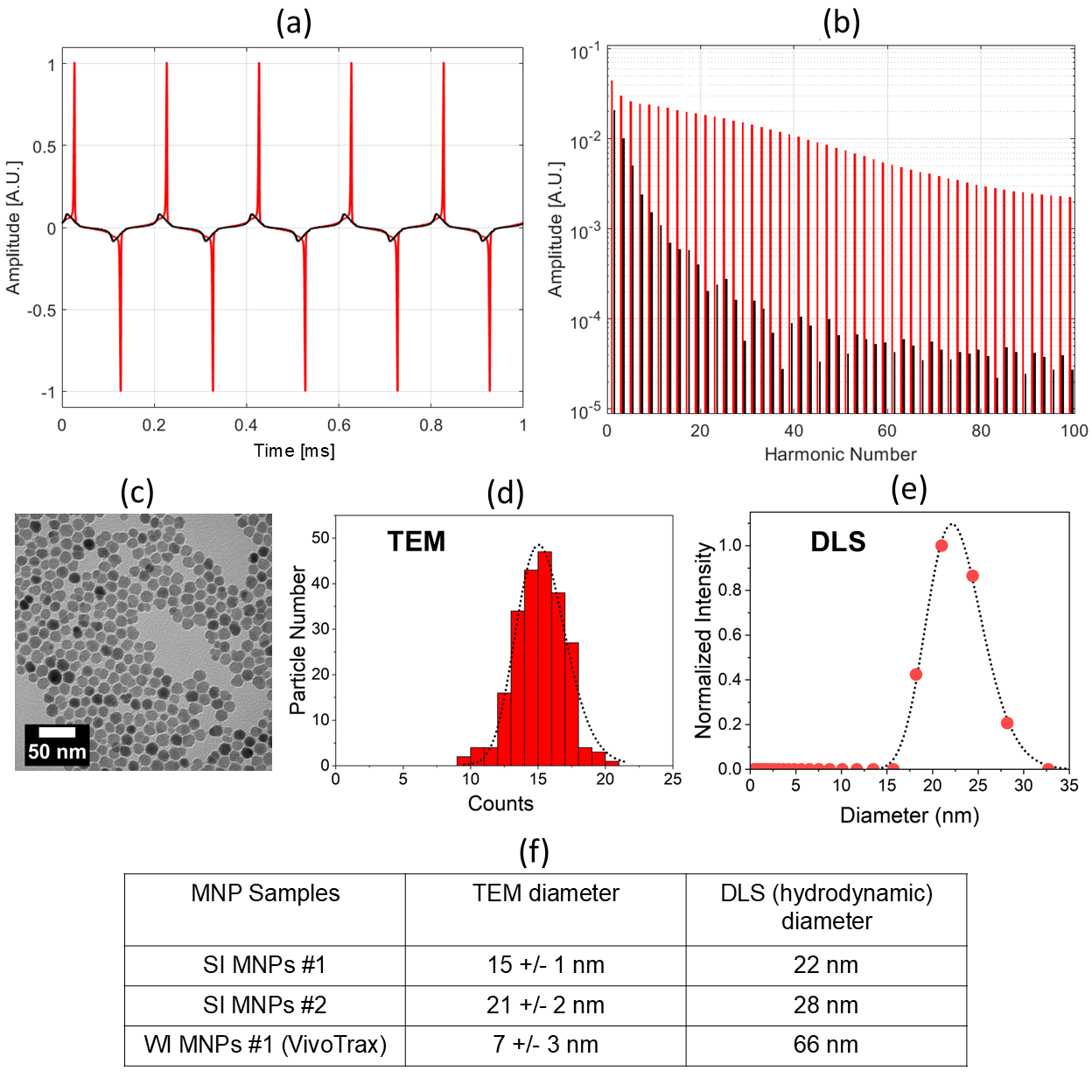}
     \captionof{figure}{AC magnetometry of strongly interacting "SI" ($\#$1, red) vs weakly interacting "WI" (Vivotrax, black) MNPs in response to a sinusoidal drive field at 10 kHz and 10 mT. (a) Voltage response amplitude (proportional to the time derivative of magnetization, Eq.~\ref{eq1}, in arbitrary units) of the SI MNPs displays sharp narrow peaks relative to WI MNPs and a phase offset. (b) Fourier transform of (a) shows a significantly broader harmonic spectrum for the SI MNPs (red) compared to the WI MNPs (black). (c) TEM images of SI MNPs $\#$1. (d) TEM diameter distribution of 100 nanoparticles for the SI MNPs $\#$1 sample. (e) Dynamic light scattering (DLS) diameter distribution of nanoparticle solution for the SI MNPs $\#$1 sample. The dotted line is a fit to a log-normal distribution.  (f) Table summarizing the size distribution of two SI MNPs systems in this work. The values for SI MNPs $\#$2  and additional physical characterization information can be found in our previous work\cite{Abel_2024}. The DLS measurements do not include a standard deviation due to the presence of multiple aggregated populations.}
     \label{fig1:v2}
 \end{figure}

% Stable colloidal dispersions of small, single-domain MNPs (or small clusters of MNPs) are typically weakly interacting (WI) with zero remanent magnetization at room temperature\cite{Kaiser_1970, Hu_2019}. These WI MNPs align with the applied external field and can assemble into larger ordered structures if the energetics are favorable: Magnetic ordering (e.g., assemblies) of MNPs is controlled by the competition between magnetic dipole-dipole interaction energy (scales as $m^2/r^3$, where $m$ is the magnetic moment and $r$ is the distance between MNP centers) and thermal energy that can break the ordering\cite{Andreu_2011}. The type of structure that is formed (e.g., circular flux closure with zero net magnetization vs. chains with a large net magnetization) is governed by Zeeman energy, which depends upon the external field $H$. Given that $m$ scales with particle volume ($V$), the interaction energy density $E_{int}/V$ scales with $R/r$, where $R$ is the MNP radius. At some critical value of $R/r$, the MNPs can be strongly interacting (SI) and display hysteretic response and remanent magnetization at zero field (sometimes referred to as superferromagnetism in literature). These ordered structures can dissipate over some timescale
% due to thermal energy agitation, resulting in reversible assemblies. \cite{
% Morup_2010, Tay_2021, Fung_2023} 
\begin{figure}
     \centering
     \includegraphics[width=17cm]{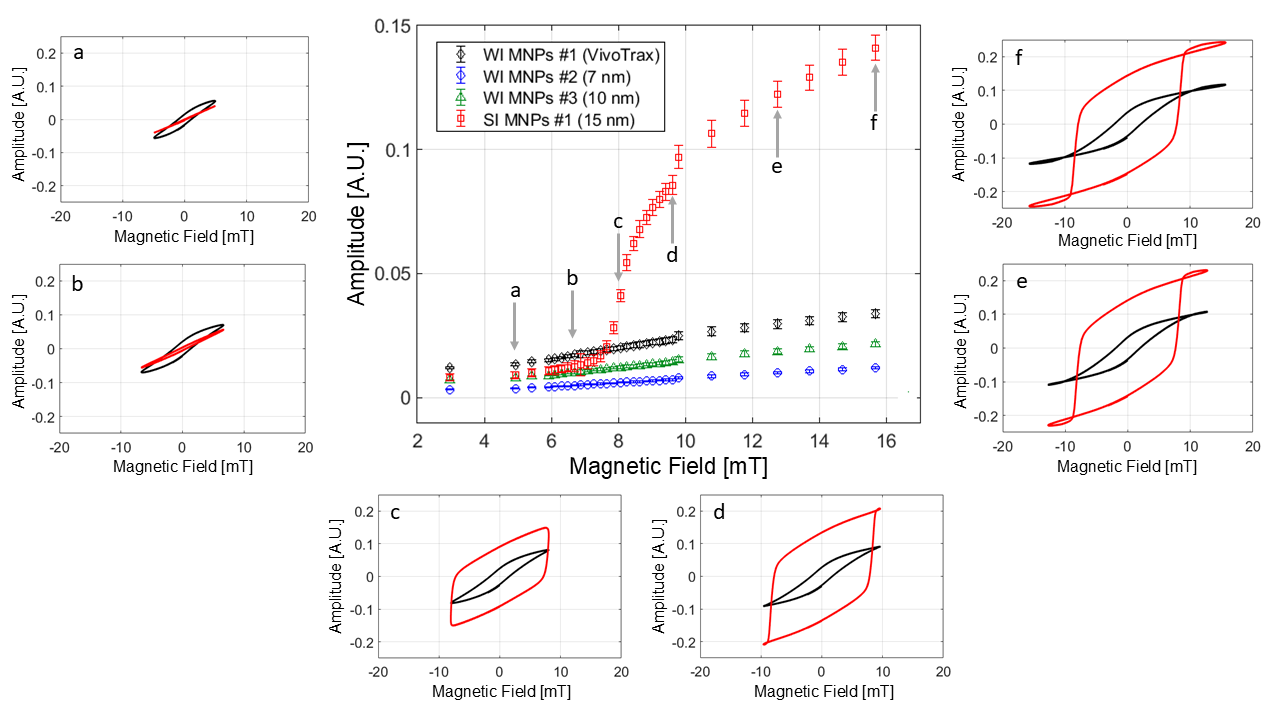}
     \captionof{figure}{Threshold behavior of SI MNPs. \textbf{Middle panel} - Voltage response amplitude (arbitrary units)  for four different MNPs at different AC field amplitudes and fixed frequency of 20 kHz. Only one (SI MNPs $\#$1, 15 nm) shows a threshold behavior marked at region "c" in the middle panel. VivoTrax (WI MNPs $\#$1) is an aggregate of single MNPs with diameter below 10 nm. \textbf{Sub panels} - comparison of the $m$ vs $H$ data for SI MNPs $\#$1 (15 nm, red) and VivoTrax (black) at the six different magnetic field points on the curve in the middle panel ("a" to "f" markers indicated by the gray arrows). The $m$ vs $H$ data are generally S-shaped for all fields for VivoTrax compared to the SI MNPs $\#$1, which shows an abrupt loop opening in the transition near the threshold field (region c).}
     \label{fig2:threshold}
 \end{figure}

We use AC magnetometry to measure different dispersions of magnetic nanoparticles. Under a sinusoidal excitation field $H(t)$, the voltage induced, $V_i(t)$, in the inductive coil sensor from the particle response is given by:
\begin{equation}
%V_i(t) \propto m\rho\dot{L}[H(t)]\label{eq1}
V_i(t) \propto m\rho\frac{d[M(H(t)]}{dt}\label{eq1}
\end{equation}
Here, $m$ is the magnetic moment, $\rho$ is the nanoparticle density (particles/volume), and $d[M(H(t))]/dt$ is the time derivative of the magnetization $M$. Figure~\ref{fig1:v2}(a) compares the measured $V_i(t)$ for the WI and SI MNPs (see Methods for details about MNP samples). In contrast to the WI system, the response from the SI system is much more sharply peaked with a distinctive phase offset\cite{Saayujya_2022, Colson_2022, Abel_2024}. The significantly broader harmonic spectrum of SI relative to WI MNPs (Fig.~\ref{fig1:v2}(b)) provides a corresponding sensitivity (order-of-magnitude voltage response enhancement of SI relative to WI MNPs) and spatial resolution boost for MPI\cite{Tay_2021}; see Fig. S1 for characterization of MPI imaging resolution for WI and SI MNPs. Figures~\ref{fig1:v2}(c)-(f) show the standard physical characterization of the main MNP samples discussed in this work. Figures~\ref{fig1:v2}(c) and (d) show a representative image of the SI MNPs and the size distribution based on TEM measurements, respectively. Figure~\ref{fig1:v2}(e) show the distribution of hydrodynamic diameters based on DLS measurements. Note that the hydrodynamic diameters from DLS are generally larger due to inclusion of the solvation layer compared to the TEM-based diameters in which the particles are dried out. The  large difference in TEM and DLS sizes for the WI MNPs $\#$1 system (Vivotrax) is due to the fact that these MNPs are $\approx$ 60 nm clusters comprised of $\approx$ 7 nm individual particles. The TEM and DLS diameters are used for estimating the Néel and Brownian relaxation time constants, respectively. Additional details for obtaining the size distributions for both TEM and DLS are found in Supplementary Figure S5. Table~\ref{fig1:v2}(f) summarizes the main SI and WI MNPs discussed in this work.    

WI and SI show distinctly different behavior when interrogated with AC magnetometry. Beyond a threshold field, the SI particles show considerable hysteresis in the $m$ vs. $H$ data.  This is exemplified in Fig.~\ref{fig2:threshold}, which plots the maximum induced voltage response as a function of $H$ and the corresponding minor loops. An abrupt transition in the minor loops is presented between subpanels b and c for the SI MNP sample. We observed a general trend that all synthesized particles of high crystal quality and whose size exceed a critical diameter of $\approx$ 15 nm\cite{Abel_2024} display threshold behavior. The two SI MNPs studied in this work show similar threshold behavior and hysteresis in the $m$ vs. $H$ data (Supplementary Figure S2). On the other hand, the smaller synthesized 7 nm (WI MNP $\#$2) and 10 nm (WI MNP $\#$3) iron oxide MNPs and VivoTrax (WI MNP $\#$1) display no threshold. These smaller MNPs are weakly interacting regardless of field strength. Finally, below the threshold field, all systems display WI behavior, indicating the threshold field is the critical field driving interparticle interactions to induce chain formation.

 \begin{figure}
  \centering
    \includegraphics[width=15cm]{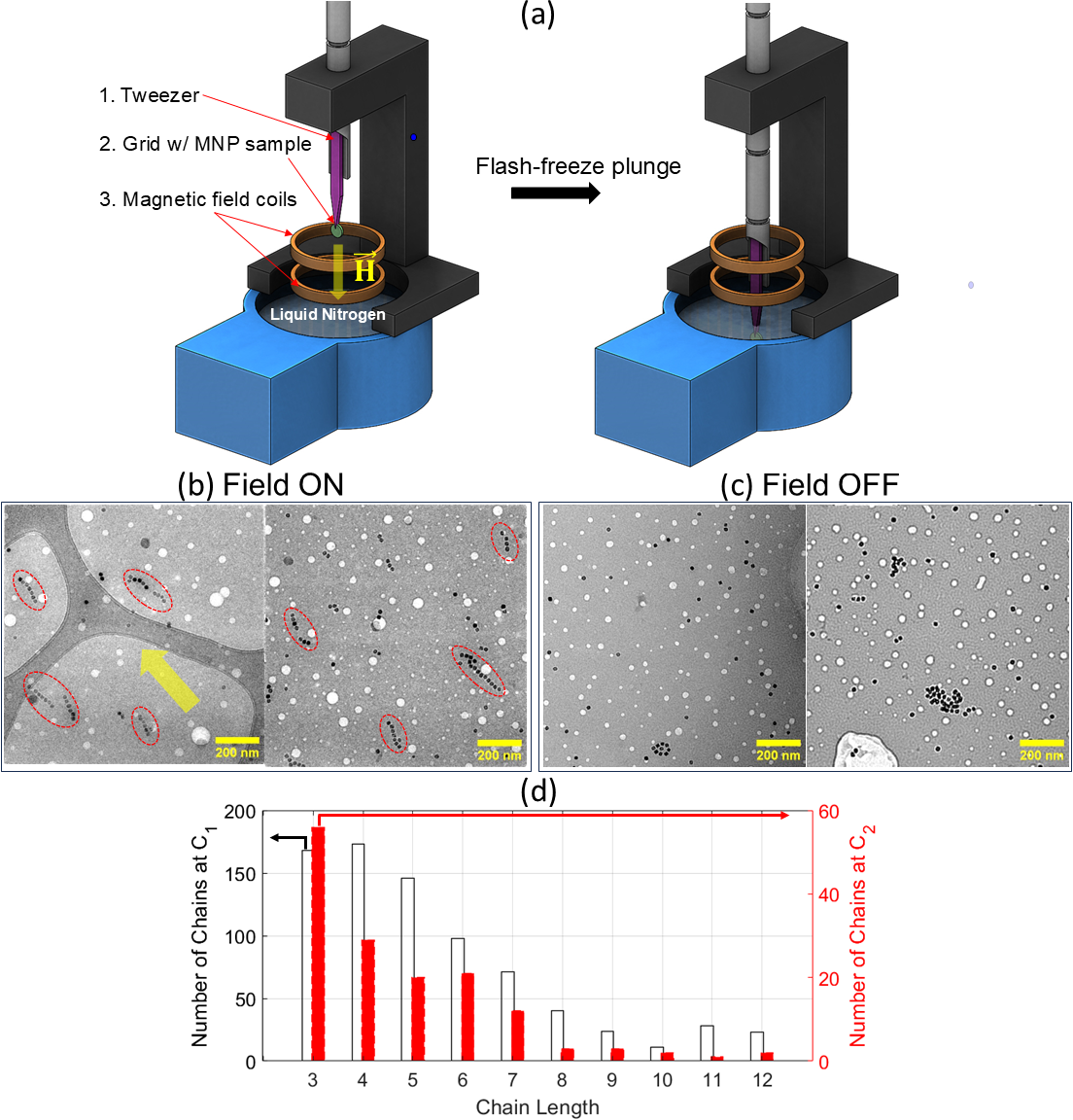}
    \captionof{figure}{(a) Cryo-TEM flash-freeze plunger method. A manual plunger equipped with a tweezer holds the sample grid (green disc) and a drop of MNPs solution (blue circle). Panels (b) and (c) show images from Cryo-TEM measurements of SI MNPs $\#$2 with the magnetic ON and OFF, respectively. For the ON case, the AC magnetic field (250 Hz, 15 mT) is turned on for about 5 seconds prior to plunging the sample grid into the liquid nitrogen bath for flash cryo-freezing. The dashed ovals indicate MNP chains composed of individual MNPs (black dots). The yellow arrow in (b) points in the magnetic field direction. Panel (d) is a histogram for the distribution of chain lengths (in units of number of MNPs) at two concentrations $C_1$ and $C_2$, where $C_1 = 2C_2$. Here, the concentrations are  $C_1 \approx 0.34$ mg/ml and $C_2 \approx 0.17$ mg/ml.}
    \label{fig3:cryoEM}
\end{figure}

\subsection*{Structure and Energetics of 1D Chains}

To reveal the physical structure that gives rise to the observed threshold behavior in our SI samples, we flash froze the MNPs in liquid nitrogen (77 K) under an AC magnetic field for observation in cryo-TEM (Fig.~\ref{fig3:cryoEM}(a)). Unlike traditional TEM where drying artifacts may be present, flash-freezing tries to preserve the sample in the same structure as its native colloidal state (in solution)\cite{Butter_2003, Taheri_2015}. Microscopy images at field on and off conditions are displayed in Figs.~\ref{fig3:cryoEM}(b) and~\ref{fig3:cryoEM}(c), respectively. Different from previous studies that used static (DC) fields with amplitudes $>$ 100 mT, our smaller field amplitude (15 mT, 250 Hz) for short duration (< 5 sec) show only the formation of isolated 1D chains with relatively short lengths, ranging from 3 to 12 nanoparticles, when compared to no field conditions. The measured edge-to-edge distance between particles in an ensemble of chains from cryo-TEM images is 2.0 +/- 1.7 nm. This is consistent with the expected non-magnetic surfactant shell thickness of $\approx$1.5 nm \cite{Oberdick_2022} from our synthesis method. We do not observe the more complex 2D or 3D assemblies described in previous studies\cite{Butter_2003,Taheri_2015,Wang_2009}, which likely formed from using a larger magnetic field and/or longer field-on time, highlighting the advantage of our method to capture the structures at these timescales. Figure~\ref{fig3:cryoEM}(d) shows the compiled histogram for the observed chain length distribution at two different concentrations. At double the concentration, the distribution is shifted to longer chain lengths with the mode at 4 MNPs per chain. 

Magnetic ordering (e.g., assemblies) of MNPs is controlled by the competition between magnetic dipole-dipole interaction energy, $E_{int}$, and thermal energy, $k_BT$ (where $k_B$ is the Boltzmann constant and $T$ is the temperature), that can break the ordering\cite{Andreu_2011,Butter_2003,Taheri_2015, Chen_2023}. The type of structure that is formed (e.g., circular flux closure with zero net magnetization vs. chains with a large net magnetization) is governed by Zeeman energy, which depends upon the external field $H$. Given that the magnetic moment, $m$, scales with particle volume ($V$), the interaction energy density $E_{int}/V$ scales with $R/r$, where $R$ is the MNP radius and $r$ is the inter-particle separation. At some critical value of $R/r$, the MNPs can form strongly interacting (SI) chains that exhibit hysteretic response and remanent magnetization at zero field. Moreover, these ordered structures can dissipate over some timescale
due to thermal energy agitation, resulting in reversible assemblies \cite{ Morup_2010, Tay_2021, Fung_2023}. 

In the following, we provide estimates for the energetics involved in our MNP systems. The interaction energy $E_{int}(\overrightarrow{r}_{ij})$ of two magnetic dipoles with dipole moments vectors, $\overrightarrow{m}_i$ and $\overrightarrow{m}_j$, separated by vector $\overrightarrow{r}$ is described by
% \begin{equation}
% E_{int}(\overrightarrow{d_{ij}}) = \frac{\mu_0}{4\pi d_{ij}^3}\Bigl[\overrightarrow{m}_i\overrightarrow{m}_j-3\frac{(\overrightarrow{m}_i\cdot{\overrightarrow{d}_{ij})(\overrightarrow{m}_j\cdot{\overrightarrow{d}_{ij}})}}{d_{ij}^2}\Bigr]\label{eq2}
% \end{equation}
\begin{equation}
E_{int}(\overrightarrow{r_{ij}}) = \frac{\mu_0}{4\pi r_{ij}^3}\Bigl[\overrightarrow{m}_i\overrightarrow{m}_j-3\frac{(\overrightarrow{m}_i\cdot{\overrightarrow{r}_{ij})(\overrightarrow{m}_j\cdot{\overrightarrow{r}_{ij}})}}{r_{ij}^2}\Bigr]\label{eq2},
\end{equation}
where $\mu_0$ is the vacuum magnetic permeability, $\overrightarrow{r}_{ij} = \overrightarrow{r}_i - \overrightarrow{r}_j$, and $r_{ij} = \mid r_{ij}\mid$. The magnetic dipole moment was estimated using 87$\%$ of bulk Fe$_3$O$_4$ ($M_s = 80$ $A \cdot m^2/kg$) to be $m$ $\approx 8\times10^4$ $\mu_B$ for $15 \pm 1$ nm diameter MNPs where $\mu_B$ is the Bohr magneton. Assuming identical particles and using a measured edge-to-edge separation of $2.0 \pm 1.7$ nm from cryo-TEM, the estimated magnetic dipole-dipole interaction energy from Eq. \ref{eq2} equates to $E_{int} = 5 \pm 3$ $k_BT$. This interaction energy that is larger than the thermal energy at room temperature is consistent with the stabilized chaining that is observed. The chaining can be sensitive to a modest magnetic field ($< 10$ mT) as well because the Zeeman energy ($\approx 2$ $k_BT$) is also on the order of the thermal energy. Since the energy density for the dipole interaction, anisotropy and Zeeman term are all invariant when particle radius ($R$) and particle separation ($r$) are scaled up in tandem (e.g., diameter and separation doubled), the energetics for chaining are the same for any system with the same ratio $R/r$.

\subsection*{Dynamics of chaining: Magnetic Relaxometry}

\begin{figure}
     \centering
    \includegraphics[width=12cm]{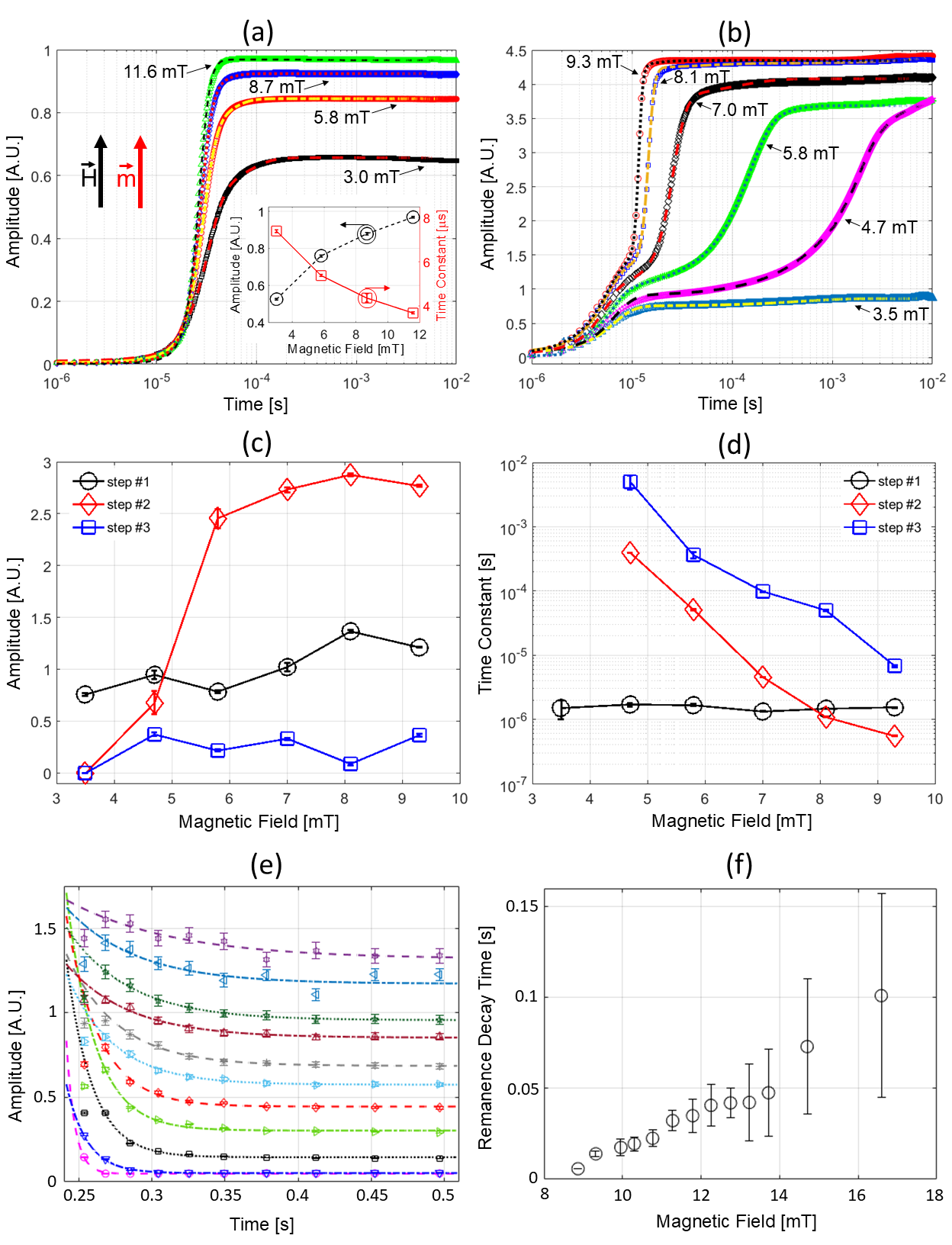}
    \captionof{figure}{MRX (pulsed) measurement. Time-dependent voltage response amplitude (arbitrary units) after turn-on of applied magnetic field of (a) WI MNPs $\#$1 (VivoTrax) and (b) SI MNPs $\#$1 at different field amplitudes. The data and corresponding fits using Eq.~\ref{eq3} are indicated by data points and lines (solid/dashed), respectively. The inset in (a) shows the fitted amplitude and time constant for WI MNPs system.  (c) and (d) show the fitted amplitudes and time constants, respectively, for the SI MNPs in (b). The solid lines in (c) and (d) are connecting lines for visual guide and not fits to the data. The "step $\#$" in the legend pertains to the three time constants in Eq.~\ref{eq3}. Step $\#$2 shows the threshold behavior for the induced voltage (proportional to the magnetic response) and displays the largest decrease in the time constants with increasing magnetic field, indicating that it is the chain formation and spin-reversal avalanche step for the entire chain. This behavior of SI MNPs contrasts with the nearly linear field-dependence for the WI MNPs displayed in the inset of (a). (e) Remanence decay (voltage response) curves after applied field is turned off for SI MNPs $\#$1 at different initially applied magnetic fields from 9 mT to 17 mT (bottom to top). (f) Fitted decay time constants as a function of magnetic field. The fitting function is an single exponential with an offset. The increased remanence lifetime with increasing amplitude indicates a memory effect.} 
    \label{fig4:MRX}
\end{figure}

Previously, Conolly and coworkers introduced a phenonemological model using a "Langevin saturator" that mimics an electronic Schmitt trigger to describe the square hysteresis loop for chain formation\cite{Tay_2021,Colson_2022,Saayujya_2022}. This mean field model captures the important idea that MNPs in a chain are strongly influenced by the combined dipole fields from neighboring particles, but it does not provide information on the coordinated process of switching along a chain. To elucidate the time dependence of magnetic switching of chains and the underlying physical mechanisms, we performed magnetic relaxometry (MRX)\cite{Bui_2022} measurements with support from micromagnetics simulations to provide direct insight on the step-by-step dynamics over a broad timescale of microseconds to seconds. Here, a fast-rise square pulse magnetic field at different amplitudes below and above the threshold field ($\approx$ 3 to 10 mT) drives the colloidal MNP systems and the steady-state response is measured. The magnetizing pulse has a rise time of of $\approx$ 1 $\mu s$, which means that magnetization dynamics on timescales slower than 1 $\mu s$ can be quantified. The time-dependent magnetization is measured by an inductive coil sensor with response time of $\approx$100 ns by integrating eq. \ref{eq1}. Figures~\ref{fig4:MRX}(a) and~\ref{fig4:MRX}(b) compare the field amplitude-dependence of response for the WI and SI systems, respectively. 

Figure~\ref{fig4:MRX}(a) shows the impulse response of the WI system for field amplitudes between 3 to 12 mT. The WI system exhibits a single exponential rise with fitted time constants between 3 and 8 $\mu$s, corresponding to a single N\'eel relaxation process. Relative to the WI system, the SI system at and above the threshold field is significantly more complex, exhibiting multistep dynamics over the five orders of magnitude in time measured. Below the threshold field ($\approx$ 5 mT), the response is characterized by a single time constant of about 10 microseconds. Above the threshold field, Fig.~\ref{fig4:MRX}(b) shows that the dynamics are dominated by two fast rises, the first around 1 to 10 microsecond (step $\#$1) and the second around 10 to 100 microseconds (step $\#$2), followed by a slow rise from 100 microsecond to seconds (step $\#$3, see Fig.~\ref{fig6:hysteresis}(c) for timescales past 1 second). As a result, an empirical, three-step sigmoidal-Boltzmann growth function was used to extract switching amplitudes (Fig.~\ref{fig4:MRX}(c)) and time constants (Fig.~\ref{fig4:MRX}(d)) from the MRX data:

\begin{equation}
\begin{split}
    M(t) 
    ={}& a_1(1-\frac{1}{1-exp((t-t_{0_1})/\tau_1)}) + a_2(1-\frac{1}{1-exp((t-t_{0_2})/\tau_2)})*\frac{1+\frac{\sqrt{(t-t_{0_2})^2}}{t-t_{0_2}}}{2} 
    \\
    & + a_3(1-\frac{1}{1-exp((t-t_{0_3})/\tau_3)})*\frac{1+\frac{\sqrt{(t-t_{0_3})^2}}{t-t_{0_3}}}{2}.
\label{eq3}
\end{split}
\end{equation}

The $\frac{1+\frac{\sqrt{(t-t_{0})^2}}{t-t_{0}}}{2}$ components in Eq. \ref{eq3} act to turn on the second and third step functions at $t > t_0$ to account for a sequential, delayed magnetization response.

The dynamic nature of the SI system's response to an impulsive excitation is evident in the observed three step magnetization that can be decomposed into three distinct physical processes as described in the following analyses. Based on an estimate of N\'eel response for isolated 15 to 20 nm MNPs in a magnetic field\cite{Deissler_2014}, step $\#$1 corresponds to N\'eel rotation of a single MNP with a time constant of about 1 $\mu$s that becomes faster with increasing magnetic field. Time constants faster than 1 $\mu$s are not quantifiable due to an instrument response limitation, and therefore, gives rise to the apparent field-independence of step $\#$1 in~\ref{fig4:MRX}(d). Note that the dynamics of the WI particles composed of small clusters\cite{Eberbeck_2013} in Fig.~\ref{fig4:MRX}(a) share a common origin as single particles of the SI system, namely a single Néel relaxation step on the timescales between 1 and 10 $\mu$s. Following the fastest (< 1 $\mu$s) dynamics, chain formation and a rapid a spin-reversal avalanche effect facilitated by dipolar interactions align the moments of neighboring MNPs in the chain over the timescale of 1 to 300 $\mu$s (step $\#$2 in \ref{fig4:MRX}(c) and \ref{fig4:MRX}(d)). Earlier work has suggested that Brownian mechanisms are the dominant form of relaxation for chain reversal\cite{Tay_2021}, but our measurements and simulations indicate that coordinated N\'eel reversal along a chain is consistent with the field-dependent time constants for magnetic switching of chains. This avalanche action can be significantly slower than the single MNP spin rotation because thermal activation causes random spin depolarization that competes directly with field alignment until sufficient time has elapsed for all MNPs in a chain to align in concert. (We provide further support for these claims in micromagnetic simulations to follow). Additional experimental evidence for the onset of chain formation and reversal avalanche is most conspicuous in response amplitude (Fig.~\ref{fig4:MRX}(c)), which shows the canonical threshold behavior (between 5 to 6 mT) that gives rise to the sharp increase in the magnetization\cite{Colson_2022}. Finally, the reversible dissipation of chains due to thermal energy when the magnetic field is removed is verified by remanence measurements using pulsed sequences similar to previous work\cite{Fung_2022,Fung_2023}. Based on this measurement (Fig.~\ref{fig4:MRX}(e)-(f)), chaining is reversible and dissipates on timescales longer than 1 ms. Since this remanence timescale is longer than the typical period of the applied AC field (100 Hz to  50 kHz), the chains do not dissipate and reform during each period of the excitation waveform.

\begin{figure}
    \centering
    \includegraphics[width=15cm]{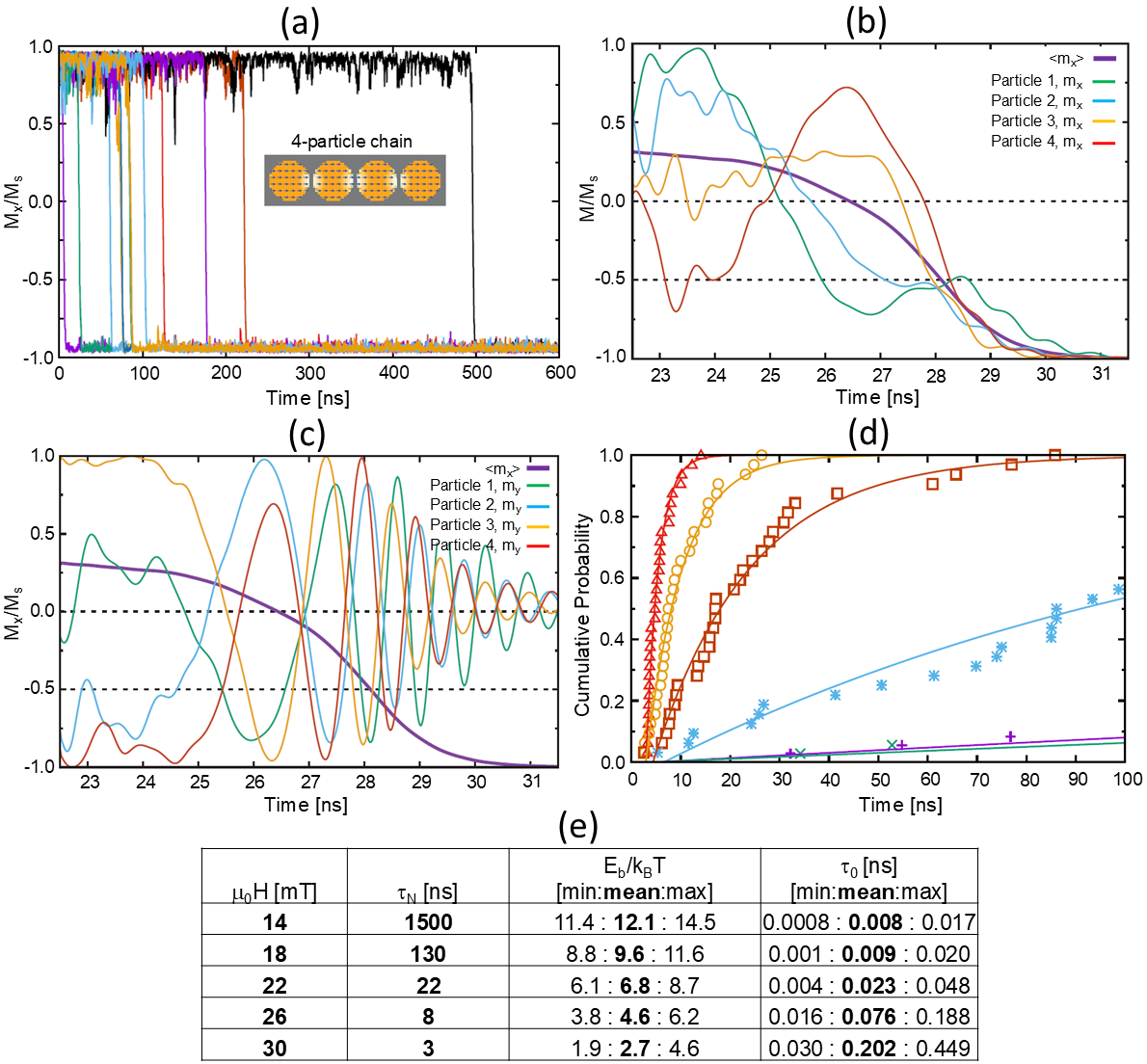}
    \captionof{figure}{Micromagnetic simulations of thermally activated N\'eel reversals for chains of four MNPs at temperature T = 290 K. Individual MNPs in the chain are 18 nm diameter spheres, with 2 nm spacing. Discretization cell size is 1 nm. Material parameters $M_s$ = 480 kA/m, A = 13.2 pJ/m, cubic anisotropy with K = -13.7 $kJ/m^3$ and damping coefficient $\alpha$ = 0.1, except as noted. (a) 12 iterations with reversing field $\mu_0 H$ = 18 mT. Applied field and chain axis parallel to $x$-axis. (b,c) Detail of a switching event with reversal field $\mu_0 H$ = 14 mT. At t = 22.5 ns the chain magnetization has just crossed over the energy barrier, and these curves show the tail of the reversal with the thermal field deactivated (i.e., T = 0 K). (b), (c) show the x- (resp. y-) magnetization component for each MNP individually (thin curves), along with the x-axis magnetization for the chain as a whole (thick purple curve). (d) Switching event times accumulated for 32 iterations at five different applied fields: $\mu_0 H$ = 30 mT (red triangles), 26 mT (gold circles), 22 mT (brown squares), 18 mT (blue asterisks), and 14 mT (green crosses and purple pluses). The curve trace marked by (+) has anisotropy $K$ = 0 $J/m^3$. Anisotropy for all others is cubic with $K$ = -13.7 $kJ/m^3$. Simulations extend to 1000 ns, but only the first 100 ns are presented for clarity. (e) Table of fitted constants for the 5 simulated iterations from (d). For each magnetic field, $\tau_N$ and the barrier height $E_b$ are fitted. The attempt time, $\tau_0$, is calculated using Eq.~\ref{eq4}. Uncertainties determined from fits to the accumulated switching event statistics in (d) using the cumulative distribution function (CDF) are provided for energy barrier and attempt time constants.} 
    \label{fig5:micromagnetic1}
\end{figure}

\subsection*{Coordinated Chain Reversal: Micromagnetic Simulation}

To understand the dynamics of this critical avalanche action in more detail, we performed micromagnetic simulations (OOMMF\cite{Donahue_1999}) of thermally-activated reversals of chains consisting of four MNPs (Fig.~\ref{fig5:micromagnetic1}). The following simulations and analysis show that the range of time constants associated with the step $\#$2 avalanche can be fully explained by coordinated N\'eel switching of nanoparticles within a chain in the presence of a thermal field. In these simulations the chain is aligned with the arbitrarily defined x-axis, and the magnetization is initialized in the +x direction. At time t = 0 ns a (pseudo-random) thermal field representing a temperature of T = 290 K is activated\cite{Palacios_1998} and a reversing field in the -x direction is applied. If the reversing field is strong enough, then after some time the thermal field can promote the magnetization over the energy barrier associated with the shape anisotropy of the chain. The time to reverse will be random, depending on the thermal field, and  Fig.~\ref{fig5:micromagnetic1}(a) shows twelve reversals at 18 mT.

Figure~\ref{fig5:micromagnetic1}(b) and~\ref{fig5:micromagnetic1}(c) provide representative details on the switching process for one iteration under a 14 mT reversing field. The thick purple curve shows the overall x-axis magnetization, while the thinner curves represent the $m_x$ and $m_y$ components for individual particles. In Fig.~\ref{fig5:micromagnetic1}(b), the thermal field initiates the reversal by nearly reversing the magnetization of the end particle (MNP 4), but its $m_x$ reverses back while the two particles at the opposite end (MNPs 1 and 2) begin their complete reversal. This is followed by MNP 3 and finally MNP 4 to complete the chain reversal. Thus, after the initial kick by MNP 4 (step $\#$1), the chain reverses in a cascading fashion from the opposite end (step $\#$2). Figure~\ref{fig5:micromagnetic1}(c) shows that this reversal is not a direct cascade, but involves precessional rotation about the chain axis as shown in the $m_y$ data. Moreover, in this rotation neighboring particles are nearly 180$^{\circ}$ out of phase, with MNP 1 and 3, and likewise MNP 2 and 4, in near alignment. This is caused by dipolar interactions between neighboring particles. A Supplementary video is included to help visualize the chain magnetization reversal process.

A cumulative distribution function (CDF) for estimating reversal time constants can be created by accumulating the switching events by time. If the switching events follow a N\'eel-Arrhenius relation,
\begin{equation}
\tau_N = \tau_0 \exp\left(\frac{E_B}{k_B T}\right)               \label{eq4}
\end{equation}
where $\tau_N$ is the mean switching time, $\tau_0$ is the attempt period, $E_B$ is the height of the energy barrier, $k_B$ is the Boltzmann constant, and $T$ is the absolute temperature, then the cumulative distribution function (CDF) has the form
\begin{equation}
\text{CDF}(t) = 1 - a^{(t-b)}                                   \label{eq5}
\end{equation}
where $b$ is a spin-up time (an offset to allow for thermal motion away from the initial magnetization configuration) and $a = 1 - 1/\tau_N$.  Thus, fitting Eq.~\ref{eq5} to curves in Fig. 4(b) allows the extraction of $\tau_N$ as
\begin{equation}
\tau_N = (1 - a)^{-1}.                                          \label{eq6}
\end{equation}
 The CDFs for several different applied fields are presented in Fig.~\ref{fig5:micromagnetic1}(d), and the fitted constants in Fig.~\ref{fig5:micromagnetic1}(e). The stronger magnetic fields in the simulations (14 to 30 mT) resulted in faster switching of chains compared to MRX experiments (step $\#$2 in Fig.~\ref{fig4:MRX}(d)), but the time constants and trend predicted by the CDFs correlate well with experimental data. Additional details on the extraction of chain switching time constants from micromagnetic simulations are provided in the Supplementary Figure S3. 

\subsection*{Chain Rotation: Experiment and Micromagnetic Simulation}

\begin{figure}
    \centering
    \includegraphics[width=15cm]{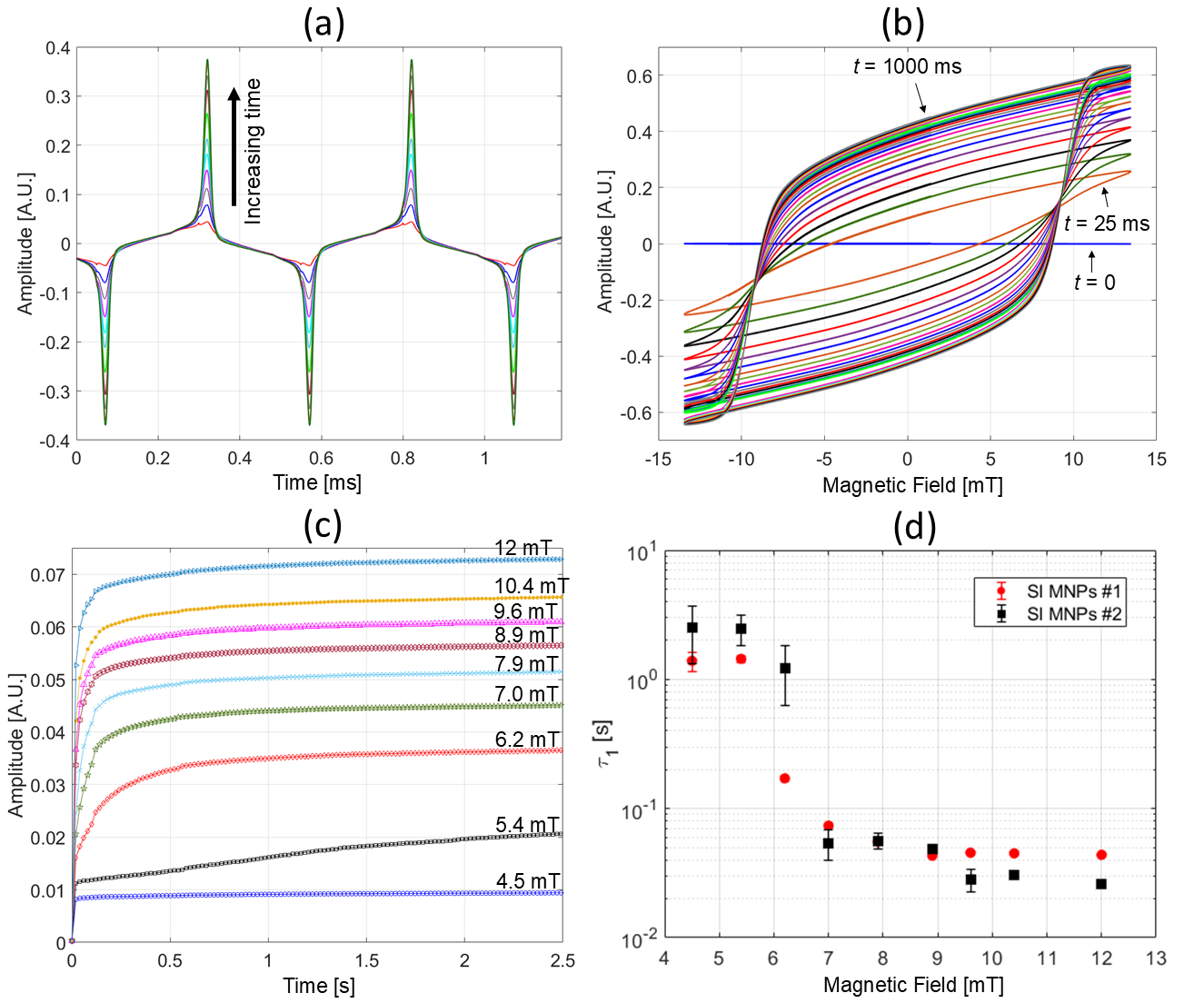}
    \captionof{figure}{Alignment of chain axis with field axis. (a) Voltage response amplitude of SI MNPs $\#$1 at 25 ms bin steps at a fixed magnetic field (10 mT, 20 kHz). As time elapses, the amplitude increases until a plateau is reached. (b) Time dependent hysteresis loops for data in (a). (c) Root-mean-square (RMS) of the voltage response amplitude for different magnetic fields at fixed frequency (20 kHz). In proximity of the threshold field (between 4.5 and 6.2 mT), the time dependence displays a clear time constant which is longer than the MRX data. (d) Fits of the fastest time constant for data in (c) to extract a time constant from the initial few data points which overlap with the MRX data show the characteristic threshold behavior. The fitted slower time constant (not shown) does not change appreciably and varies between 300 ms to 1 s.}
    \label{fig6:hysteresis}
\end{figure}

To probe the origin of the third slow time constant (100s of microseconds to seconds, step $\#$3) in the MRX data, an experiment was performed where a sinusoidal magnetic field was turned on at time zero and the magnetization response was continuously recorded and binned in 25 ms intervals (total time of 2 s). This measurement tracks the  temporal evolution of the $m$ vs. $H$ data. Figure~\ref{fig6:hysteresis}(a) shows that the SI MNPs response increases over time and does not peak within one cycle of the AC excitation field, as was also observed by Colson {\it{et al.}}\cite{Colson_2022};  this can also be seen in the  corresponding evolution of the $m$ vs. $H$ data (increasing coercivity and remanence over time), plotted in Fig.~\ref{fig6:hysteresis}(b). Using data from Fig.~\ref{fig6:hysteresis}(a), Fig.~\ref{fig6:hysteresis}(c) shows the calculated RMS response over time, and reveals that the evolution timescale depends on the magnetic field strength. Here, the first point (amplitude) for each measurement trace is not zero, suggesting that fast dynamics prior to 25 ms has occurred. Fits to capture the fastest time constant exhibited by data in~\ref{fig6:hysteresis}(c) are presented in Fig.~\ref{fig6:hysteresis}(d), and once again show the canonical threshold behavior, indicating that chains are responsible for the dynamics. 

\begin{figure}
    \centering
    \includegraphics[width=15cm]{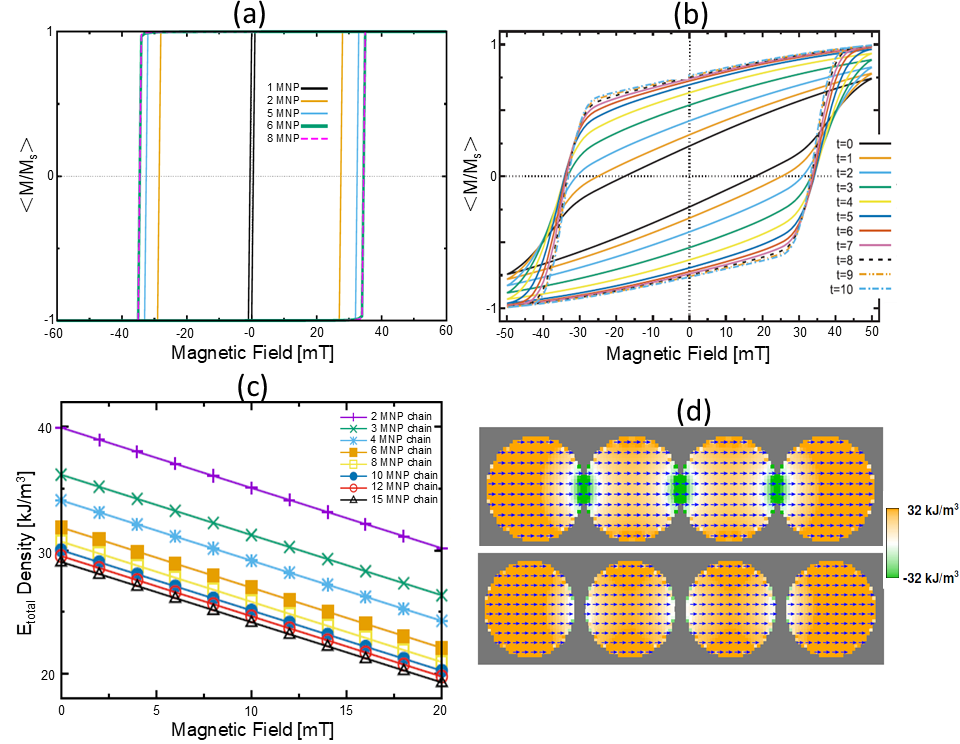}
    \captionof{figure}{Micromagnetic simulations comparing chain growth vs chain rotation mechanisms: (a) Chain growth: $m$ vs $H$ data with increasing chain length. The coercivity increases rapidly from monomer to dimer, but saturates with increasing particle count. (b) The cumulative response for a collection of 200 4-MNP chains that are initially randomly oriented in 3D (time t=0, arbitrary units). The chains gradually rotate into alignment with the applied field axis at a rate proportional to the torque on the chain (i.e., $sin(\theta)$ where $\theta$ is the angle between chain and field axis). The hysteresis loops become increasingly square-like and saturate by t=10 when the chains are all nearly aligned with the applied field axis. This result captures the essential features of the experimental data in Fig.~\ref{fig6:hysteresis}(b). (c) Magnetic energy density as a function of an on-axis applied field for chains with various number of 18 nm MNPs, with fixed 2 nm spacing between the MNPs. (d) Spatial variation in energy density for two 4 nanoparticle chains with no applied field. The nanoparticles are 30~nm across, with 0~nm and 4~nm spacing in the upper and lower figure, respectively.  The interior particles have overall lower energy (green regions) than those on the ends due to magnetostatic interaction with the particles on either side, explaining the reduced energy density for longer chains seen in Fig. 7(c).}
    \label{fig7:micromagnetic2}
\end{figure}

We used full 3D micromagnetic modeling to explain the dynamics and energetics of these chains. Most interestingly, the gradual growth in the remanence and coercivity over time in Fig.~\ref{fig6:hysteresis}(b) is suggestive of a Brownian alignment of the chain axis to an external field, rather than chain growth: the latter mechanism would result in an abrupt increase in the remanence and coercivity between the transition from monomer to dimer and saturate rapidly with increasing chain length (see simulations in Fig.~\ref{fig7:micromagnetic2}(a)). To model the evolution of magnetization from chain rotation, the $m$ vs. $H$ data for individual 4-MNP chains at various angles to an applied field (fixed orientation) are computed (see Methods section for details). Fig.~\ref{fig7:micromagnetic2}(b) shows the cumulative response for a collection of 4-MNP chains that are initially randomly oriented in 3D (time t=0, arbitrary units). As the chains gradually rotate, the hysteresis loops become increasingly square-like and saturate by t=10 when the chains are all nearly aligned with the applied field axis. This result captures all the essential features of the experimental data in Fig.~\ref{fig6:hysteresis}(b). Note that thermal effects could explain the observed lower coercivity in the experimental data compared to simulations. Our analysis indicates that the physical mechanism responsible for the final slow increase of magnetization in step $\#$3 is coordinated rotation of the chain axes toward the field direction.

Finally, micromagnetics simulations can shed light on the observed chain length distribution from cryo-TEM data. Fig.~\ref{fig7:micromagnetic2}(c) presents the magnetic energy density for chains of various lengths as a function of an on-axis applied field. At all fields the overall energy is less for longer chains, so it is energetically favorable for short chains to lengthen as indicated in the following analysis. At a fixed field (e.g., 6 mT), a 4-particle chain has an energy density of $\approx 32$ $kJ/m^3$, corresponding to a total energy of $3.9\times10^{-19}$ $J$. An 8-particle chain has a lower energy density ($\approx 28$ $kJ/m^3$), but due to its larger volume, its total energy is $6.8\times10^{-19}$ $J$. Importantly, two 4-particle chains have a combined energy of $7.8\times10^{-19}$ $J$, which is higher than that of a single 8-particle chain. However, this energy reduction saturates beyond six MNPs, allowing other forces in the system to limit chain length. This is consistent with mode of the chain length in the population distribution being 4 MNPs or less with very few chains longer than 8 MNPs. This trend is further explained in Fig.~\ref{fig7:micromagnetic2}(d) (cross section through the center of the particles in the full 3D simulation), which shows that interior particles experience lower energy due to magnetostatic interactions with neighboring particles. Longer chains contain a higher fraction of interior particles, thus reducing the average energy per particle. While Fig.~\ref{fig7:micromagnetic2}(d) illustrates this effect for 30 nm particles at zero field, the same principle applies to 18 nm particles shown in Fig.~\ref{fig7:micromagnetic2}(c), where the magnetization is nearly uniform. Under an applied field aligned with the magnetization, the configuration remains mostly unchanged, and the primary field-dependent contribution is the Zeeman term, which varies linearly with field strength, as observed in Fig.~\ref{fig7:micromagnetic2}(c). Here, the energy decrease afforded by nanoparticle chaining is caused primarily by a local reduction in the magnetostatic energy near neighboring nanoparticles. The 32 $kJ/m^3$ energy density volumes in the center of the nanoparticles (orange) is mostly self-magnetostatic (i.e. shape anisotropy) energy from the individual MNPs. The lower energy regions (-32 $kJ/m^3$, green) represent inter-particle stray field interactions. This latter effect quickly dissipates with increased spacing between the MNPs (upper vs. lower figure).

\subsection*{Conclusion}

This work addresses the physical structure and step-by-step magnetization dynamics of strongly interacting MNP dispersions driven by AC magnetic fields. Previous literature has noted the enhanced response and coercivity of these systems compared to their weakly-interacting counterparts, but evidence for the structure and physical mechanisms under the field drive conditions of MPI has been limited. Using cryo-TEM, we provide direct evidence of the physical structure as simple 1D chains that are responsible for the complex magnetization dynamics and ascribe each observed step measured by MRX to a physical relaxation mechanism (single particle Néel, chained Néel, and Brownian rotation of chains). These strongly-interacting magnetic nanoparticles that display 1D chaining have already been harnessed to enhance the performance of magnetic particle imaging (MPI) tracers, but to date a detailed physical understanding of their time-dependent switching has been lacking\cite{Tay_2021,Abel_2024}. The field-driven hysteresis may also be useful for hyperthermia applications, where a large hysteresis loop is desirable for optimal heat generation \cite{Tay_2021,Nemati_2016,Mille_2021,Fung_2023,Niculaes_2017}. Looking forward, we envision the prospect of developing nanoscale vehicles that can robustly stabilize these chains in a solid matrix for advancing magnetic nanoparticle-based sensing. Our micromagnetic modeling indicates that the microsecond-range chain switching dynamics are captured by N\'eel mechanisms, which will be preserved in solid systems. From a computational perspective, these colloidal dynamics from relatively simple MNP assemblies could aid to advance all-atom simulations to account for complex contributions from the solvent and ligand composition of the shell on magnetodynamics\cite{Mahmood_2021}, critical for applications that use MNPs including soft robotics, directed drug delivery, and biological sensing. Finally, the presented dynamics of interacting MNPs could be fundamental for the understanding of phase transitions in dipolar fluids\cite{Tlusty_2000}. 

\newpage

\section*{Methods} 

\subsection*{Magnetic nanoparticle synthesis and physical characterization}

Two methods were used for synthesis of colloidal nanoparticles: Protocol A was used for synthesis of the three samples with diameters 7 nm (WI MNPs $\#$2), 10 nm (WI MNPs $\#$3), 15 nm (SI MNPs $\#$1) and Protocol B was used for synthesis of the 21 nm (SI MNPs $\#$2) sample. 

Protocol A is a modified procedure of Sun et al. to synthesize iron oxide nanoparticles via thermal decomposition\cite{Sun_2002}. Each synthesis was modified slightly to produce nanoparticles with different sizes. Walter Stern 501 carbon microporous boiling chips were used in the synthesis.

The smallest particles (7 nm, WI MNPs $\#$2) were synthesized by mixing 1.70 g oleic acid, 1.60 g oleylamine, 0.71 g iron acetylacetonate and 2.58 g of 1,2 hexadecanediol with 15 mL of benzyl ether in a 125 mL volume 3-neck flask. The mixture was attached to Schlenk line and degassed three times with nitrogen until a pressure of 25 mTorr was achieved. For the reaction, the solution was raised to a temperature of 200° C for 2 hours. Then, the solution was heated to a reflux temperature of 298° C for 1.5 hours. During the reaction, the solution was stirred continuously using a magnetic stir bar at 1,150 RPM. After the reaction, the solution was cooled to room temperature and washed 2x with a 1:1 mixture of isopropyl alcohol and acetone. 

 WI MNPs $\#$3 (10 nm) was synthesized by mixing by mixing 0.95 g oleic acid, 0.90 g oleylamine and 0.95 g iron acetylacetonate with 10 mL of benzyl ether in a 125 mL volume 3-neck flask. For this reaction, a magnetic stir bar was not used for mixing during the reaction since larger nanoparticles are attracted to the stir bar. Instead, 0.6 mL of boiling chips were added to the reaction vessel. The mixture was attached to Schlenk line and degassed three times with nitrogen until a pressure of 25 mTorr was achieved. For the reaction, the solution was raised rapidly a reflux temperature of 298° C for 1 hour. After the reaction, the solution was cooled to room temperature and washed 2x with a 1:1 mixture of isopropyl alcohol and acetone. 

SI MNPs $\#$1 (15 nm) was synthesized using the same procedure as Sample B, except with an increased mass of each precursor. For this reaction, the precursor masses were 1.21 g oleic acid, 1.14 g oleylamine and 1.2 g iron acetylacetonate. All other conditions were kept the same. 

TEM Preparation and Imaging: To prepare samples for TEM, the stock nanoparticle solution was diluted by a factor of 1/50 to 1/100 in toluene. Then, $\approx$ 20 $\mu L$ of the dilute solution was pipetted on top of a Formvar/carbon coated copper TEM grids (Ted Pella). Imaging was performed using a Tecnai T12 Spirit BT transmission electron microscope with a LaB$_6$ filament. 

Protocol B (SI MNPs $\#$2) was developed to produce magnetic nanoparticles with similar SI MNPs properties. Synthesis and characterization details are published in recent work\cite{Abel_2024}. Briefly this method is similarly based on the thermal decomposition of iron acetylacetonate, using 10 mmol of 1,2 hexadecanediol to produce high crystal quality of particles\cite{Moya_2015,Nedelkoski_2017}, in a mixture of 1-octadecene (primary solvent) and oleic acid (surface functionalizing ligand). The sized was tuned by changing the heating rate to produce particles in the desirable size regime. 

The crystallographic quality of iron oxide nanoparticles (and corresponding magnetization) for different synthesis types prepared using Protocol A and B formed single crystals\cite{Abel_2024, Nedelkoski_2017}. These particles also have high magnetization – close to bulk Fe$_3$O$_4$. High resolution TEM in both studies show the synthesis formed single crystals. 

The mass magnetization (moment per unit mass of magnetite) of the MNP samples was measured using a combination of magnetometry (to measure magnetic moment) and inductively coupled plasma optical emission spectroscopy (to measure the mass of iron). The magnetite mass was calculated from the measured iron mass using the assumption that the particles had magnetite (Fe$_3$O$_4$) stoichiometry. The measured particle magnetization at 300 K ranged from 77$\%$ to 97$\%$ of the magnetization for bulk magnetite (92 Am$^2$/kg). 

Additional details about the determination of nanoparticle core and hydrodynamic diameters using TEM and DLS methods can be found in the Supplementary Figure S5. 

\subsection*{AC Magnetometry and Magnetic particle imaging (MPI)}

Details of the high speed AC (oscillating and pulsed) magnetometry instrumentation are reported previously\cite{Bui_2022}. Briefly, a custom arbitrary-wave magnetic particle spectrometer was used for characterizing MNPs (magnetic particle spectroscopy and hysteresis) over broad frequency (100 Hz to 50 kHz) and amplitude ($<$ 50 mT) using both sinusoidal and pulsed fields. The excitation solenoid coil and detection inductive coil are made from low inductance Litz wire. The inductive coil sensor is in a gradiometer configuration. For pulsed field measurements, electronics set a lower limit for a measurable time constant at $\approx$ 1 $\mu s$. 

A custom magnetic nanoparticle imaging (MPI) instrument\cite{Bui_2023} was used to image the phantom of MNP solutions displayed in the Supplementary Figure S1. The excitation AC magnetic field ($f_{excitation}$ = 31.75 $kHz$, 10 $mT_p$) and a spatial selection, gradient magnetic field ($H_{gradient}$ = 20 $T/m$ in the x-axis, and 10 $T/m$ in the y-axis and z-axis) were used for the imaging. The gradient magnetic field was generated from a pair of N52 grade NdFeB permanent magnets. 

\subsection*{Cryo-TEM}

Particles suspended in toluene at room temperature were applied to grids, followed by application of the magnetic field, blotting off excess solution, and vitrification by plunge-freezing. Sample concentrations were diluted by 10 to 100 fold from the native concentration to avoid completely attenuating the electron beam. The SI MNPs behavior is preserved at these dilute concentrations (Supplementary Figure S4). Grids were ultrathin continuous carbon on lacey carbon supports (Ted Pella \#01824G). Freezing was performed using a manual plunger \cite{nguyen_manual_2022} with the grid plunged directly into liquid nitrogen to induce flash-freezing. The liquid nitrogen Dewar was filled to near the top, and the solenoid was suspended roughly 1 cm above the liquid nitrogen surface. A grid was clamped in tweezers and suspended at the midpoint of the solenoid using the plunging apparatus, then 3 $\mu l$ of ferrofluid was applied to the grid using a pipette. The magnetic field (250 Hz, 15 mT) was then applied, the grid blotted manually for 5 seconds using a thin strip of filter paper while still inserted in the solenoid, and the grid plunged into liquid nitrogen immediately. The 5 seconds field-on time is longer than needed for chains to align to the field, consistent with a uniform alignment of the chains in a particular direction in the cryo-TEM data.

Electron microscopy was performed using a Glacios CryoTEM instrument (Thermo Fisher Scientific) operated at 200 kV and a nominal magnification of 36,000$\times$. Imaging was performed with a Ceta detector yielding a calibrated pixel size of 0.412 nm. SerialEM\cite{mastronarde_sem_2005} was used for automated data collection. Exposures were collected at an electron dose rate of 2360 $nm^{-2} s^{-1}$, using 5 frames of 0.5 seconds each for a total exposure of 2.5 seconds and total electron dose of 5890 $nm^{-2}$. Motion-correction was performed using the native implementation within \mbox{Relion}\cite{nakane_relion_2018}.

\subsection*{Micromagnetics simulation (OOMMF)}

The micromagnetic simulations were performed using the OOMMF micromagnetic package from NIST \cite{Donahue_1999}. Except as otherwise noted, the MNPs were modeled as 18 nm diameter spherical particles with saturation magnetization $M_s$ = 480 $kA/m$, exchange constant $A$ = 13.2 $pJ/m$, magnetocrystalline anisotropy constant $K$ = -13.7 $kJ/m^3$ (easy axis aligned to chain long axis), and Gilbert damping parameter $\alpha$ = 0.1. Discretization cell size was set at 1 nm, well below the exchange length $l_{ex} = \sqrt{2A/(\mu_0 M_s^2)}$ = 9.5 nm typically recommended as the upper limit for discretization.  Quasi-static computations for hysteresis loop simulations were performed via conjugate-gradient energy minimization, and thermal studies used a $2^{nd}$ order Heun method to integrate the stochastic Landau-Lifshitz-Gilbert equation with thermal field of the Langevin type. \cite{Palacios_1998} The particle orientations are held fixed in all the micromagnetic simulations, so magnetization reversals are necessarily of the N\'eel type.

Chain rotation simulation: To model the rotation of a collection of MNP chains, hysteresis loops for individual 4-MNP chains at various angles to an applied field are computed. Geometric and compositional differences between chains are allowed by averaging each loop against copies of itself scaled to have coercive fields normally distributed with relative variance 0.01.  The response of a collection of chains with varying orientations can then be computed by summing hysteresis loops corresponding to the individual orientations. The evolution of an ensemble of chains are modeled by gradually rotating each chain in proportion to the torque on it from the applied field.  The field step size is 1 mT. Preliminary runs with larger field step sizes produced similar results, so we believe the results are converged at 1 mT. The magnetic response from a background population of isolated MNP particles are included to provide a non-hysteretic upward cant (slope) to the response.

\section*{Data availability}

The datasets used and/or analyzed during the current study are available from the corresponding author on reasonable request.

\section*{Acknowledgments}

We thank Angela Hight Walker and Kevin Douglass for helpful discussions. We acknowledge financial support from the NIST Innovation in Measurement Science (IMS) program. S.D.O. performed TEM at the University of Colorado, Boulder using the EM Services Core Facility in the MCDB Department. S.D.O. acknowledges funding from the National Science Foundation (NSF) (https://ror.org/021nxhr62) through NSF-CBET award no. 2038046. S.D.O also acknowledges funding from PREP agreement no. 70NANB18H006 between the National Institute of Standards and Technology (NIST) (http://ror.org/05xpvk416) and the University of Colorado, Boulder (http://ror.org/02ttsq026).

Reference is made to commercial products to adequately specify the experimental procedures involved.  Such identification does not imply recommendation or endorsement by the National Institute of Standards and Technology, nor does it imply that these products are the best for the purpose specified. 

%\bibliography{chaining_BIB}

\end{document}